# An In-Depth Exploration of the Effect of 2D/3D Views and Controller Types on First Person Shooter Games in Virtual Reality

Diego Monteiro  
Xi'an Jiaotong-Liverpool University

Hai-Ning Liang[1]  
Xi'an Jiaotong-Liverpool University

Jialin Wang  
Xi'an Jiaotong-Liverpool University

Hao Chen  
Xi'an Jiaotong-Liverpool University

Nilufar Baghaei  
Massey University

**ABSTRACT**

The amount of interest in Virtual Reality (VR) research has significantly increased over the past few years, both in academia and industry. The release of commercial VR Head-Mounted Displays (HMDs) has been a major contributing factor. However, there is still much to be learned, especially how views and input techniques, as well as their interaction, affect the VR experience. There is little work done on First-Person Shooter (FPS) games in VR, and those few studies have focused on a single aspect of VR FPS. They either focused on the view, e.g., comparing VR to a typical 2D display or on the controller types. To the best of our knowledge, there are no studies investigating variations of 2D/3D views in HMDs, controller types, and their interactions. As such, it is challenging to distinguish findings related to the controller type from those related to the view. If a study does not control for the input method and finds that 2D displays lead to higher performance than VR, we cannot generalize the results because of the confounding variables. To understand their interaction, we propose to analyze in more depth, whether it is the view (2D vs. 3D) or the way it is controlled that gives the platforms their respective advantages. To study the effects of the 2D/3D views, we created a 2D visual technique, PlaneFrame, that was applied inside the VR headset. Our results show that the controller type can have a significant positive impact on performance, immersion, and simulator sickness when associated with a 2D view. They further our understanding of the interactions that controllers and views have and demonstrate that comparisons are highly dependent on how both factors go together. Further, through a series of three experiments, we developed a technique that can lead to a substantial performance, a good level of immersion, and can minimize the level of simulator sickness.

**Keywords**: Virtual Reality, 2D/3D Views, Controller types, First Person Shooter, Gaming, Head-Mounted Displays.

**Index Terms**: [Human-centered Computing]: Human-Computer interaction—Interaction Devices; [Human-centered Computing]: Human-Computer interaction—Interaction paradigms—Virtual reality;

## 1 INTRODUCTION

Virtual reality (VR) has been proliferating rapidly in the last few years, especially with the advent of mass commercial VR Head-Mounted Displays (HMDs). Due to their rapid growth, there are still aspects that are not well understood, including the influence of different types of controllers in virtual interactions [1]–[4], and how they affect immersion [5]. Similarly, the effects of viewing types on Simulator Sickness (SS) during fast-paced applications, such as First-Person Shooter (FPS) games [6]–[9], are also underexplored.

Players pay attention to several factors during gameplay [10], [11]. In VR, the level of playability and immersion are important considerations because they affect enjoyment and performance, especially while players are (or are not) feeling SS symptoms. Playability and immersion can be affected by how the VR environment is controlled [12]–[14]. Nevertheless, we must note that research shows that immersive technologies can hinder performance when compared to a traditional setup (i.e., monitor, keyboard, and mouse) [15], [16]. From these studies, it is not clear if it is the keyboard and mouse combination as the input control responsible for the positive results. Likewise, it is not possible to know whether the enhanced results are the product of the reduced depth or visual detail of the 2D display that allows FPS players to perform better (e.g., having a better aim). After all, real-life marksmen close one eye when shooting, which effectively renders their vision to 2D mode as a filtering mechanism. Finally, it is not clear if better performance is caused by the combination of visual reduction of details and how players control or navigate the environment.

In this paper, we set up to investigate the effect of 2D and 3D views and the controller types in immersive VR for FPS gameplay. To this end, our aim is to analyze: (1) the effect of variations of those views on immersion and performance in VR; (2) the impact of using a standard keyboard and mouse combination compared to hand-held controllers that are used in today's VR systems.

To study the effects of the 2D views, we created a 2D visual technique, which we call PlaneFrame. It is a technique that slightly alters how the VR environment is viewed by users. Through three experiments, we found that this technique can improve performance when playing an FPS game in VR, and decrease the level of SS with little effect on the users' perceived level of immersion. The results of our studies show that PlaneFrame could be a useful visual technique for FPS games and other VR applications.

Our main contributions in this paper are a deeper understanding of the impact of 2D/3D views on VR FPS and the influence traditional input methods have on these kinds of VR applications. Further, we found that a 2D view in VR HMDs can be a good compromise between performance and immersion and can help reduce symptoms of SS. Finally, we introduced PlaneFrame. This versatile and cost-effective visual technique can be used for VR applications, including FPS games and exploratory environments.

## 2 RELATED WORK

Our research is related and informed by three main themes: (1) View Modality; (2) Simulator Sickness; and (3) Gameplay Performance. First, we discuss viewing technologies and how they affect immersion and enjoyment. Then, we summarize previous studies on mitigating SS associated with a higher level of immersion. Last, we present previous studies that have dealt with the trade-off between immersion and performance.

---

[1] Corresponding author (haining.liang@xjtlu.edu.cn)

It is common to separate the terms immersion from presence. However, due to their close connection, they are often used interchangeably [17]. Hence, we also adopt this approach.

## 2.1 2D/3D Views in VR

While commercial VR HMDs are relatively new, techniques that focus on increasing the sense of immersion are not. One such example is commercial 3D TVs, which have been used to demonstrate that having a 3D image changes how information is processed in the brain. They can elicit not only a greater sense of presence but also greater vection (the sensation of movement of the body in space produced purely by visual stimulation) when compared to 2D images [18]. Such findings are consistent with research that has evaluated 2D/3D VR when comparing a 3D wall display against a 2D monitor [19]. This research seems to indicate that people process 2D and 3D views differently. However, both of these studies were conducted in CAVE like displays to navigate mazes. Hence, it is still unclear if the same results would apply to other tasks, like looking for fast-moving opponents in a game, or whether more immersive technologies like VR HMDs [20] would produce the same results.

One trait that seems consistent among the different studies is that people often seem to yield the highest enjoyment from the most immersive experiences, especially in HMD type of displays [15], [16], [20]–[24]. This is the case even though HMDs tend to cause nausea and other kinds of simulator sickness afflictions [18], [22]. Most of these studies compared HMD VR, PC, or a CAVE in some combination. However, to our knowledge, there has not been a study that has compared 2D and 3D views within an HMD. As such, it is unclear whether the higher level of satisfaction comes from the HMD display, regardless of whether it shows 2D or 3D views.

## 2.2 Simulator Sickness

Simulator sickness (SS) or kinetosis has been studied even before the arrival of VR HMDs [25]. Because it is often believed that SS is caused by the brain receiving conflicting information from the senses, some researchers assume that the closer we get to the real physical environment, the less likely a person is to feel the symptoms of SS [26]. Although some researchers have suggested that such an approach is adequate [27] or at the very least indifferent [28], Dziuda et al. [29] have re-evaluated this assumption to identify some sort of uncanny valley of SS. They showed that a simulator which included a moving platform associated with the visual stimulus generated higher levels of SS. Their results would imply that trying to go for the highest possible realism might not be an effective way to prevent SS. Because the actual mechanisms for SS might be different from those of motion sickness, an earlier study found it hard to correlate motion sickness with SS [30].

One technique that is simple but has yielded great success is the use of a gazing point in front of the user's view. The gazing point can be a circular point [31] or the head of a character in the case of the third-person perspective (3PP) view [23], [32]. Further, this is somewhat a diegetic technique and can be easily inserted as the gun's aim and can be quite useful. The counterpoint is that gazing cues can attract the player's eyes [33], which could affect users' ability to see objects that fall outside of the area of the gazing point and are located in their peripheral vision.

More advanced and intricate techniques, such as galvanic vestibular stimulation, require extra hardware and might not be usable by all users (e.g., users wearing pacemakers cannot use them) [34], [35]. Other techniques that seem to yield promising results are the ones that use vibration to stimulate the sense of movement, such as bone-inducted vibration [36] or producing small strikes on the region behind the ears [37]. Nevertheless, these techniques are still in their early stages, needing further development and evaluation to assess their real effectiveness, suitability, and long-term effects.

Recently, techniques that are developed primarily for VR HMDs (rather than all kinds of simulators) have been proposed. These techniques focus more on how virtual worlds are presented rather than the movement or navigation technique alone. For example, some have attempted to apply blur during movement according to the distance of objects to the user [38]. Others deformed the peripheric region in which the user is moving towards to create the illusion of less movement or reduced FOV [39], [40]. Other techniques involve removing visual information from the user [41], [42]. While these techniques seem to present positive results in lowering SS, it is unclear if these optical illusions affect user performance and their perceived level of immersion in the VR environment. To the best of our knowledge, only two other studies have evaluated the effect of their mitigation technique on performance [43], [44]. They involve mainly changing the rotation aspect, either reducing the speed during rotation or removing information. Their techniques interfered little with the user's performance in their respective applications. As such, it is likely that mitigation techniques are a promising path for VR games, especially for FPS games that require fast head movements [44].

## 2.3 FPS Performance

FPS games are a well-established genre, with reasonably simple gameplay mechanics and are well suited to be translated into VR due to its inherently first-person view format. Unavoidably, there are challenges to bring an established format (in 2D displays) to a new platform like VR. One possible problem is the loss of player performance. For example, in a study that compared a CAVE to a PC FPS, participants performed considerably worse in CAVE; however, the same participants declared having enjoyed more the CAVE experience [15], [16]. It could be likely that VR HMDs may suffer from the same issue, trading immersion for performance [45]. However, not all studies agree with this view; [17], when trying to replicate a study that compared an HMD and a desktop display for a Role-Playing Game (RPG) [46], found no significant differences between two in terms of user satisfaction and presence. One reason for the discrepancy is because FPS is naturally more engaging. A second explanation might be the kind of controller that players used, namely keyboard and mouse in all platforms, which could potentially break the immersion.

One issue with these studies is the type of controller used because the studies had their controllers match the technology (i.e., keyboard for the desktop and a game controller for VR). As such, their results might have had more to do with the controller than the display. After all, earlier studies on the influence of the type of controller had found a difference in presence when comparing a more natural Wiimote to a PlayStation game controller for a tennis game [5]. There are even differences in brainwaves when comparing both controllers [47].

For FPS games, [4] found that there is no difference between the Oculus Touch dual hand controller and the Xbox game controller [4]. However, because the Oculus Touch is more closely related to the Xbox controller than to the keyboard and mouse, it may not be possible still to discard the possibility that the type of controller could affect immersion and performance [48]. In another study, it was found that the keyboard and mouse led to lower performance [3]. Like other studies, the authors also matched the controller to the technology. With new proposed designs emerging for VR FPS controllers [13], the search for a suitable controller is an on-going process [45] and an important aspect to be explored to seek ways to improve both immersion and performance for FPS VR games.

In short, prior research shows that the closer we get to simulating a real environment, the lower the performance can get. Also, it is not clear how this affects SS. This research aims to fill this gap by investigating the following research questions.

*RQ1.* Is performance loss caused by the VR environment or by the type of controller used?
*RQ2.* Do visual stimuli play a more prominent role in in-game subjective feelings than how we control them?
*RQ3.* Is the stereoscopic 3D view a significant factor within the HMDs?

To explore these three questions, we devised three studies. To our knowledge, there was not a technique or tool that allows rendering a 2D view inside the Stereoscopic display, like VR. Thus, we developed one inhouse. In the next session, we present this technique, which creates a customizable 2D Frame, which we dubbed PlaneFrame.

## 3 PLANEFRAME – 2D IN VR

To study the effect of a 2D screen like view for HMDs for FPS VR games on performance, level of immersion, and degree of control, we developed PlaneFrame (PF). PF displays a copy of the 3D virtual environment (VE) within a user's field-of-view (FoV) through a 2D rectangular plane (see Figures 1 and 2). When navigating in the VE, the 2D plane appears in front of the user in a fixed distance (see Figure 2a). The user can perceive the 3D VE through the 2D plane, like looking at 3D views through a 2D display. The main difference between PF and the gazing point technique [26] is that PF carries a 2D copy of the 3D VE region at which the user is looking. As such, with PF, there is no blind spot or hidden peripheral regions. This technique also differs from Slave Visualization [41], which is a technique not aimed at changing the whole VR environment but at sharing 3D views among VR users. As we show in Experiment C, the size of PF is configurable. It also works as a fixed gazing point to help avoid or minimize SS.

To make PF easily configurable and fit diegetically in different kinds of application scenarios, we developed two variations, called Mask and 2D-3D Extreme Mask.

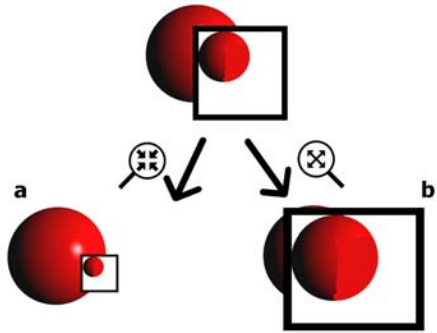

Figure 1: The concept used in the PlaneFrame technique and the configurable size of its FoV. (a) A smaller-sized 'mask.' (b) A bigger-sized mask.

### 3.1 Mask View

A 'Mask' can have an area (A) that occupies between 2.5% and 100% of the FoV (height × width) of the HMD (see Figures 1 and 2). Mask is a rectangle plane in which the center is in position (0,0), with a given height (H) and width (W). We define the PF FoV ($P_{FoV}$) to be based on H and W, as given in equations 1 to 4, where $s$ is the desired size of PF.

$$k = \sqrt{s}, \ 0.025 < s < 1 \quad (1)$$
$$H = height \times k \quad (2)$$
$$W = width \times k \quad (3)$$
$$P_{FoV} = H \times W \quad (4)$$

### 3.2 2D-3D Extreme Mask

2D-3D Extreme Mask View (or simply Extreme Mask), like the Mask, would allow the user to see the area in front of him/her and converted the 3D stereoscopic view into a 2D view. The size of PF was also configurable. However, in this version, the user can only see the area within the Mask, and the 3D VE is hidden (see Figure 2c and 2d). One main difference of Extreme Mask is that it was continuously activated, whereas, in the other versions, it was only activated during motion. This could lead to a more immersive experience because it would remove all external distractions and changes in view.

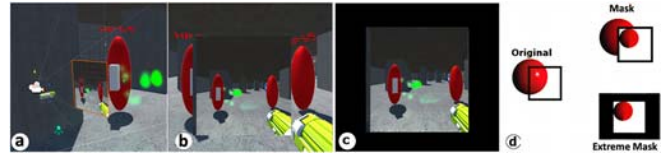

Figure 2: (a) PlaneFrame (PF) is a technique that displays a copy of the 3D virtual environment (VE) on a rectangular plane which is placed in front of the camera (or users' view). (b) A demonstration of the PF Mask version used in a first-person shooter (FPS) game. (c) A demonstration of the PF Extreme Mask version used in the game. (d) Illustrations of the concept of Mask and Extreme Mask: they both display a screenshot of the 3D VE within a user's field-of-view through a 2D rectangular plane. Mask keeps the 3D VE behind the plane, whereas Extreme Mask hides the 3D VE completely.

## 4 GAME ENVIRONMENT AND EXPERIMENT METRICS

We conducted three experiments to understand how the PF would affect our participants' level of immersion, SS, and gameplay performance. Experiment A collected keyboard data and compared different views, including unmodified VR 3D view and regular 2D monitor. In Experiment B, we compared PF with another existing technique popular in FPS games, namely "Gazing point" [31]. Experiment C evaluated the performance of three variations of the PF to help us understand in more depth the effect of different size of PF on induced SS, user performance, and immersion. Experiments A and B are also used to understand the effects of controller types (i.e., keyboard and controller) for FPS games in VR from a between-subjects perspective.

The FPS game used as a testbed for our experiments was developed inhouse. This allowed us to avoid or minimize any confounding factors and to implement the different versions of PF and game to fit the specific controllers and displays. In the game, players have to navigate rapidly while performing other actions typical of an FPS game [42].

Next, we describe the game environment.

### 4.1 Game Environment

The goal of the game was for players to traverse a maze with hallways, open spaces, and chambers and try to survive and destroy (kill) easily distinguishable non-player characters (which we called adversaries). There were three types of adversaries with three different behaviors. One adversary, Sniper, was stationary. The second, dubbed Patroller, could only walk along a fixed route. The third adversary, the Hunters, could hunt the player (based on an AI algorithm provided in [49]). To make them easily distinguishable, they were presented in three different colors: green, blue, and red (see Figure 3). The adversaries were positioned throughout the maze, progressively increasing in numbers until the final chamber (see Figure 4).

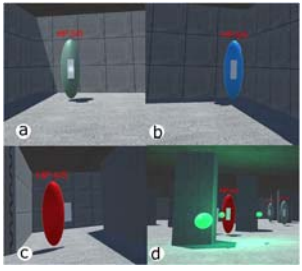

Figure 3: Three kinds of adversaries or enemies present in the game environment: (a) green adversary, a Patroller; (b) blue adversary, a Sniper; (c) red adversary, a Hunter; (d) a player is inside the main chamber of the maze which contains several adversaries of each kind.

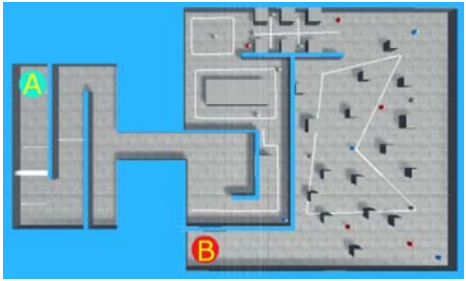

Figure 4: Bird' s-eye view of the maze used in the experiment. A player must eliminate all adversaries going from point A to point B (the final chamber).

The maze was designed with tall gray walls and turning points, which forced the player to move in a non-straight path. This design slowly introduced the adversaries to the players. The lack of visual cues was meant to reduce path memorization, and the turns were meant to assess the occurrence of SS, as the act of rotating around the x-axis tends to increase the levels of SS during gameplay [36], [50]. This kind of turning also allows us to verify the accuracy of the different controller types (see next section). Like any FPS game, there were many positions where the player could hide, but to finish the game, the player had to reach the end of the maze (see Figure 4). There were no ambiguous paths or splits; thus, all the players would follow roughly the same path.

### 4.1.1 Evaluation Metrics

For all the experiments, we use the Simulator Sickness Questionnaire (SSQ) [25], and Immersive Experience Questionnaire (IEQ) [26]. The SSQ contains three parts, measuring the level of Nausea, Dizziness, and Oculomotor issues. These three parts are combined to give the overall weighted level for SS (Total Severity). Each symptom uses a 5-point Likert Scale, ranging from least severe to most severe. Each symptom might count points towards more than one part.

The scoring system for the game was as follows: when a player gets hit, he or she loses 10 points; when a player hits a target, he or she gets 10 points. Negative punctuation is possible. Based on this system, we define Accuracy (Acc) as the number of shots that hit the target divided by the number of shots performed by the player. The formula for Accuracy can be found in equation 6, where $i$ is the total number of shots and $j$ is the shots that did not find a target

$$Acc = \frac{i-j}{i} \quad (6)$$

Even though a player could be killed, the player's health points (HP) was set to be able to endure at least 4 minutes of straight-shooting. This life span guaranteed that the players could advance through the maze long enough to allow collecting meaningful results. The IEQ also uses a 5-point Likert Scale for each question. In the end, all questions are summed.

## 5 EXPERIMENT A

Experiment A was designed to (1) compare how an immersive environment compares to a regular computer screen; (2) determine the influence of a 2D view and; (3) collect the gameplay data for when players use the keyboard only. In short, we compared induced SS and user performance with or without PF when playing an FPS game in VR HMD systems. We also included a desktop version as another comparative baseline. Thus, this experiment evaluated three versions: conventional PC display (CD), HMD VR display without PF (RVR), and HMD VR display with PF (VRPF). For this experiment, we only used Extreme Mask.

### 5.1 Participants

We recruited 18 participants (6 females; 12 males) from a local university. They had an average age of 18.89 (s.e. = 0.99), ranging between 17 and 21. All volunteers had a normal or normal-to-corrected vision, and none of them declared any history of color blindness or other health issues, physical or otherwise. Sixteen participants (88.9%) had experience with VR systems before the experiment. Three participants (11.1%) declared having already felt a certain degree of sickness when playing FPS games in the common types of display (laptop, desktop, or TV).

### 5.2 Apparatus

We used an Oculus Rift CV1 as our HMD, as it is one of the most popular off-the-shelf VR devices. The HMD was connected to a desktop with 16GB RAM, an Intel Core i7-7700k CPU @ 4.20GHz, a GeForce GTX 1080Ti dedicated GPU, and a standard 21.5" 4K monitor. We used a mechanical keyboard and high-resolution gaming mouse as input instead of the traditional Oculus Touch, which is one factor we are collecting data in this experiment.

In the VR versions, although participants could not see the keyboard, they were familiar with using the keyboard and mouse when playing FPS games. Nevertheless, to diminish the likelihood of it being a confounding factor, they were not required to move their fingers to different keys on the keyboard after the initial positioning. Further, it is essential to note that the position of the monitor was calculated to emulate the size of the VRPF; this would eliminate FOV as a confounding factor. Our VRPF configuration in Unity was (90,180,0) for rotation, (0.1, 0.1, 0.1) for scale, and (0, 0.687, 1.663) for position.

### 5.3 Experimental Procedure

On arrival, each participant was assigned a specific order of the three versions in which he or she would play the game. The order was counterbalanced with a Latin Square design approach to mitigate carry-over effects. The participants then filled in a pre-experiment questionnaire that collected demographic and past gaming experience information.

After that, participants were presented with a demonstration of the Oculus Rift and a virtual environment without walking to get them acquainted with the VR HMD and setup. Next, we introduced the input device and the rules of the game to each participant.

Participants played the three versions in a pre-defined order and, after completing each version, were asked to answer the questionnaires mentioned above. The participants were required to rest (and as much as they wanted) before they could go to the next version.

## 5.4 Results

The data were analyzed using both statistical inference methods and data visualizations. We conducted a Shapiro-Wilk test to check the normality of the data. For those that were classified as normal, we used parametric tests; for the others, we used non-parametric ones.

For normally distributed data, we conducted Mauchly's Test of Sphericity. We also employed Repeated Measures ANOVA (RM-ANOVA) using Bonferroni correction to detect significant main effects. If the assumption of sphericity was violated, we used the Greenhouse-Geisser correction to adjust the degrees-of-freedom. Partial Eta-Squared is declared for the ones with significant main effects. We conducted Friedman tests for non-normal data. When there was a detectable significance, we ran separate Wilcoxon signed-rank tests on the different combinations to pinpoint where the differences occur.

To make a consistent presentation of the plots, in all of them, we used separate colors and patterns to represent the three versions: CD (blue-checked), RVR (red-diagonally stripped), and VRPF (green-plain). The data were presented with outliers for a deeper understanding of the results.

### 5.4.1 Simulator Sickness Questionnaire (SSQ)

Figure 5 summarizes the results of the SSQ data and shows that RVR presented the most significant change in the level of SS that the other versions. The data did not follow a normal distribution. The Friedman test showed that there was a statistically significant difference in Oculomotor, Dizziness, and Total Severity (hereafter Total) based on which version the game was being played, $\chi^2(2) = 11.815$, $p = .003$, $\chi^2(2) = 16.618$, $p = .000$, and $\chi^2(2) = 16.395$, $p = .000$ respectively. Nausea did not appear to differentiate significantly according to the test $\chi^2(2) = .323$, $p = .851$.

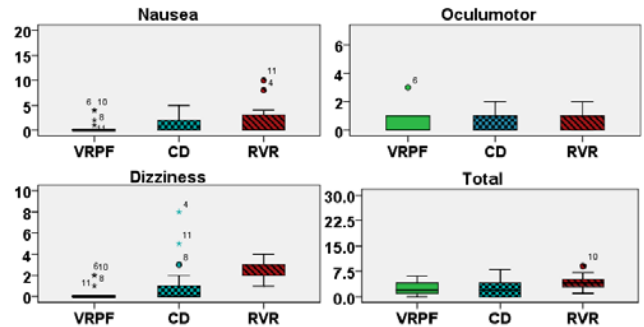

Figure 5: The results of SSQ based on the four sub-scales in Experiment A. RVR presented the most significant change in the level of SS compared to the other versions. VRPF and CD did not differ significantly between themselves.

During our post-hoc analysis with Wilcoxon signed-rank tests, we observed the Median (IQR) Oculomotor levels for CD, VRPF, and VR were 0, 0, and 1, respectively. After correction there were no significant differences between them, except between RVR and CD ($Z = -2.333$, $p = 0.020$). The lack of significance is clear between CD and VRPF ($Z = 0.000$ $p = 1.000$), but less so between RVR and VRPF ($Z = -1.684$, $p = 0.092$).

The post hoc analysis further showed significant differences in Dizziness between VRPF and RVR ($Z = -3.299$, $p = 0.001$). However, there was no significant difference between VRPF and CD ($Z = -1.476$, $p = 0.140$) or between RVR and CD ($Z = -1.764$, $p = 0.078$). Their Median (IQR) were CD = 0, RVR = 2 and VRPF = 0.

What is important to note from this analysis is that Total diverged significantly between RVR and CD ($Z = -2.759$, $p = 0.006$), and between RVR and VRPF ($Z = -2.831$, $p = 0.005$). The other pair did not have significant difference ($Z = -.224$, $p = .823$). 2, 4, and 2 were the Median IQR of CD, VRPF, and VR, respectively.

### 5.4.2 User Performance

Figure 6 shows the data for Score, Time, and Accuracy across the three versions. The Shapiro-Wilk test revealed that the Score measurements did not follow a normal distribution, but Time and Accuracy did. The RM-ANOVA suggested that there were significant main differences in Accuracy ($F2, 34 = 7.304$, $p = .002$, $\eta p2 = .313$) among the three versions. The post-hoc analysis showed that Accuracy in FPS-VR was significantly lower than the other two versions (p=.013 and p=0.14). No other significant relations were found for Time ($F2, 34 = 1.448$, $p = .250$); on average each condition lasted about 251 seconds (4.2 minutes).

The Friedman test revealed no significant difference for Score $\chi^2(2) = 3.254$, $p = 0.197$.

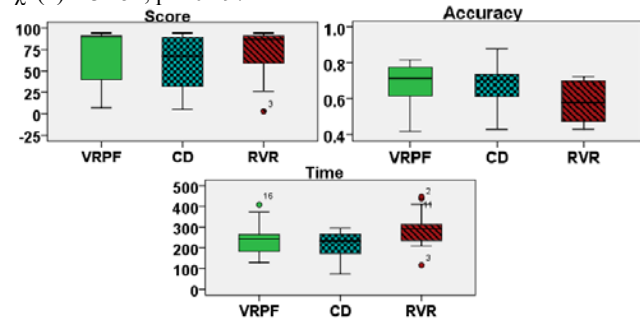

Figure 6: The results of Score, Accuracy, and Time. Accuracy in RVR was significantly lower than the other two versions.

### 5.4.3 Subjective Immersion

Figure 7 shows the data for Overall Immersion across the three versions. The RM-ANOVA suggested that there were significant main differences ($F2, 32 = 3.852$, $p = .032$, $\eta p2 = .194$) among the three versions. Post-hoc analyses using Tukey's HSD indicated that Overall Immersion was greater on VRPF when compared to VR ($p = .030$).

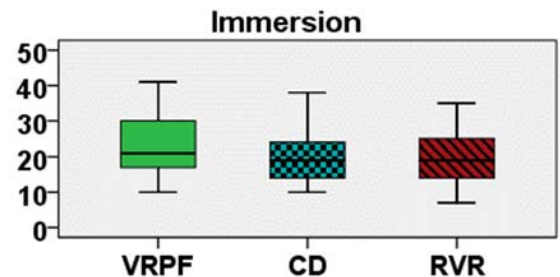

Figure 7: The results of Overall Immersion. Overall, Immersion did differ significantly in VRPF compared to the other versions.

## 5.5 Discussion

The results show that PF was a promising approach to lower the level of SS for FPS VR games. Results from SSQ showed that, with PF, participants had significantly lower levels of Oculomotor, Dizziness, and Total Severity than playing in the original, unmodified VR environment. It seemed that by projecting the 3D VE onto the 2D replica, we could decrease players' level of SS significantly while allowing them to navigate in the VE and did not lead them to lose the context of the environment. Our results also showed that when using PF, the VR environment might lead to similarly low levels of SS, as does the CD. This offers the possibility that users could play or, more generally, navigate in VEs with a low level of SS in VR systems with PF.

Concerning user performance, the results indicated that PF led to a significantly higher accuracy than the original VR version and had similar performance with the PC version. We could infer that because, with a lower SS, the players were able to shoot at the targets more accurately in the FPS VR game. This result is significant because this could mean that the lower SS could lead to a more precise view of the elements of the environment, and this could have applications outside of FPS games.

Further, what was surprising was that VRPF was statically more immersive than RVR. We had expected that the free nature of RVR to be completely immersive and allow players to forget they were in a VE. Based on the results, we were able to identify two hypotheses for why RVR suffered in this experiment:

*HA1*. Playing with the keyboard did not allow the users to be fully immersed in the experience.

*HA2*. Given the total lack of other stimuli in the VRPF, the players got solely focused on the game and thus felt immersed.

In summary, when compared to a regular FPS in VR, PF presented positive results in the reduction of SS and supported users to achieve high accuracy. Since the standard display is currently the baseline for most FPS games, we can infer that PF represents a viable technique to reduce SS in VR while allowing users to achieve performances that are comparable to the current best environment.

## 6   EXPERIMENT B

Experiment B aimed to compare PlaneFrame (PF) and Regular FPS VR with a diegetic technique that can potentially improve simulator sickness and increase Accuracy during FPS gameplay in VR HMD systems: gaze point [31]. Three versions were compared: HMD VR display without PF (RVR), HMD VR display with PF (VRPF), and Gaze, which gave the users a point to stare at continuously (see Figure 8). Our Gaze point was a circular green ring (with a radius of 15 Unity Units) located at the center of the camera. It was permanently on the screen and followed the player's head movement. The Gaze point was not used as a game controller.

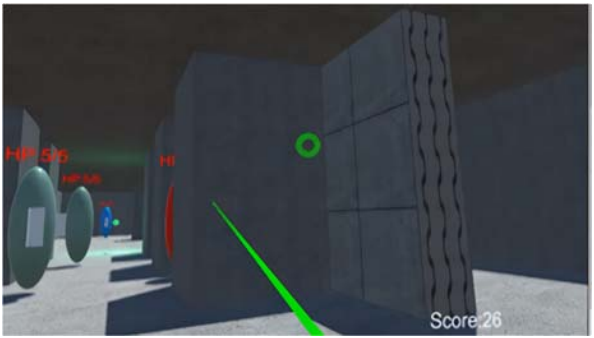

Figure 8: A screenshot during gameplay using the Gaze technique which provides a gazing point represented by the green ring. The player could stare at the point during navigation to mitigate SS effect and potentially improve aim.

### 6.1   Participants, Apparatus, and Experimental Procedure

Another 18 participants (equally divided between males and females) were recruited from a local university with an average age of 20.0 (s.e. = 1.90) ranging from 18 to 27. All volunteers had a normal or normal-to-corrected vision, and none declared any history of color blindness or health issues. Thirteen participants (72.2%) had experience with VR systems. Two (11.1%) reported feeling sick during regular FPS gameplay. We used the same VR device and the desktop PC configuration as in the first experiment.

Participants also followed a similar procedure to complete the three versions offered in this experiment.

### 6.2   Results

We employed a similar procedure to analyze the data as the first experiment. We used the following colors to represent the three versions in the plots: Gaze (blue-checked), RVR (red-diagonally stripped), and VRPF (green-plain).

#### 6.2.1   Simulator Sickness Questionnaire (SSQ)

Given that the SSQ data were not normally distributed, we performed a series of Friedman tests, which showed that there was a statistically significant difference in all sub-scales and Total (see Table 1). The results of the SSQ in terms of sub-scales and Total are summarized in Figure 9.

Table 1.  Summarized results of Friedman tests of SSQ, significant differences in all subscales.

| Scale | $\chi^2(2)$ | p |
|---|---|---|
| Nausea | 16.618 | .000 |
| Oculomotor | 16.745 | .000 |
| Dizziness | 11.577 | .003 |
| Total | 11.789 | .003 |

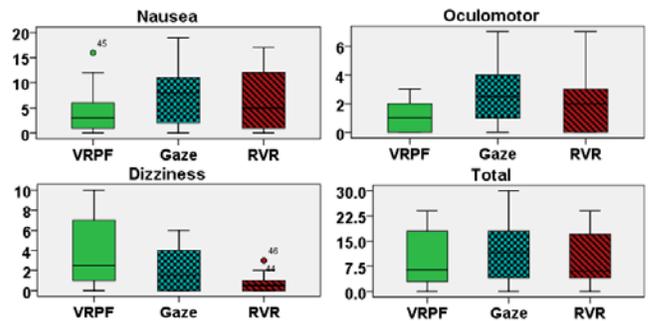

Figure 9: The results of SSQ concerning the four sub-scales of Experiment B. VRPF presented positive results in Nausea and Oculomotor but underperformed in Dizziness.

The post-hoc analysis with Wilcoxon signed-rank tests showed that the Median (IQR) Nausea levels for the Gaze, VRPF, and RVR were 8, 3, and 5, respectively. After correction there were significant differences between VRPF and Gaze (Z = -3.428, p = .001) and VRPF and RVR (Z = -1.960, p = .050), but not between Gaze and RVR (Z = -.828, p = .408).

Post-hoc analysis of the Oculomotor component was similar to that of Nausea, revealing significant differences between VRPF and Gaze (Z = -3.325, p = .001) and VRPF and RVR (Z = -2.183, p = .029), but not between Gaze and RVR (Z = -1.592, p = .111). Median (IQR) Oculomotor levels for the Gaze, VRPF, and RVR were 2.5, 1, and 2.

Interestingly, all comparisons presented significance in the post-hoc analysis of the Dizziness component: VRPF and Gaze (Z = -1.968, p = 0.049) and Gaze and RVR (Z = -2.365, p = 0.018). The strongest effect was between RVR and VRPF (Z = -3.051, p = 0.002). Median (IQR) Dizziness levels for the Gaze, VRPF, and RVR were 1.5, 2.5, and 0.5.

Finally, our post-hoc comparison showed that Total diverged significantly between Gaze and VRPF (Z = -3.190, p = .001). the other pair combinations did not have significant difference: RVR Gaze (Z = -1.709, p = 0.087) and VRPF and VR (Z = -1.156, p

=.248). The Median IQR were Gaze = 11.5, RVR = 6 and VRPF= 6.5 respectively.

### 6.2.2 User Performance

The RM-ANOVA indicated that there were no statistically significant differences in Accuracy ($F_{2,34} = 2.605$, $p = .089$) or Time ($F_{2,34} = .493$, $p < .615$); on average each version lasted 280 seconds (4.6 minutes). However, the Friedman test revealed significant differences for Score $\chi^2(2) = 9.662$, $p = 0.008$. Figure 10 shows the results.

Post-hoc analysis revealed that the Score in VRPF was significantly lower than RVR ($Z = -2.369$, $p = .018$) and Gaze ($Z = -3.247$, $p = .001$), while the Median (IQR) Score levels for the Gaze, VRPF and RVR were 76.5, 60.5 and 74.5. Gaze and RVR did not present significant differences ($Z = -0.327$, $p = .744$).

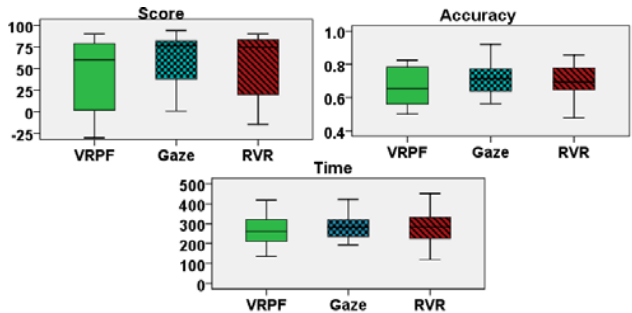

Figure 10: The results of Score, Accuracy, and Time for Experiment B. VRPF performed worse than the two other versions in Score.

### 6.2.3 Subjective Immersion

Figure 11 shows the data for Overall Immersion across the three versions. The RM-ANOVA showed that there were significant main differences in Overall Immersion ($F_{1.508, 25.635} = 7.849$, $p = .004$) among the three versions. Post-hoc analysis showed that Overall Immersion in Gaze was significantly higher than in the VRPF ($p=.04$). VRPF and RVR ($p=.02$) also presented significant differences. Gaze and RVR ($p>.05$) did not present significant differences.

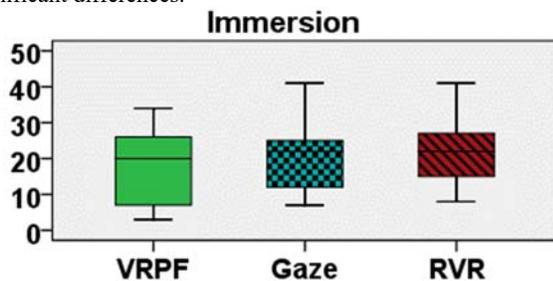

Figure 11: The results of Overall Immersion in Experiment B. VRPF presented significant differences compared to the two other versions.

### 6.3 Discussion

From our analysis, we observed a clear difference among the three versions regarding induced SS. The Total Severity was significantly different between VRPF and Gaze, but not between VRPF and RVR. These results are somewhat surprising. We expected Gaze to perform better than no technique at all. VRPF performed visibly better in most subscales except in dizziness. The results suggest that the FP 2D view helps to mitigate much of the symptoms of simulator sickness but can also cause disorientation, which should be investigated further (see Experiment C, in the next section).

The counterpoint to this distinctly lower level of simulator sickness is the lower levels of immersion generated. This result is especially interesting since, in Experiment A, we observed VRPF generating the highest levels of Immersion. This divergence might be caused by the influence of the controller. According to the results, if the users use a keyboard, they will feel more immersed in a 2D view while, when using the controller, using a 3D view will likely increase their level of immersion.

This effect extends further as VRPF has led to lower Score results and can have implications on the controller aspect as well. The results together seem to imply that the type of controller could affect the speed of navigation, which in turn could have a disorienting effect when looking at 2D and 3D views. The effect could be enhanced further in the latter. Disorientation could invariably affect performance and lower scores.

## 7 EXPERIMENT C

The goal of this experiment was to compare different variations of PF regarding induced SS, user performance, and immersion when playing an FPS game in VR HMD systems only. We varied the size of PF to 30% (SM) and 95% (BM) of the HMD's FoV. We were interested in exploring if the size of PF would influence the dependent variables. Therefore, in this experiment, we compared three versions in total: VRPF, SM, and BM.

### 7.1 Participants, Apparatus, and Experimental Procedure

We recruited another 18 participants (five females, twelve males, and one non-binary) from a local university with an average age of 19.7 (s.e. = 1.73) ranging from 18 to 24. All volunteers had a normal or normal-to-corrected vision, and none declared any history of color blindness or health issues. Thirteen participants (72.2%) had experience with VR systems. Three (16.7%) declared feeling sick during FPS gaming sessions with regular displays. We used the same VR device, and the desktop PC as in Experiment A. Participants followed a similar procedure as the other experiments to complete the three versions for this experiment.

### 7.2 Results

We employed a similar procedure to analyze and represent the data as the first experiment. We used the following colors and patterns to represent the three versions in the plots: BM (blue-checked), SM (red-diagonally stripped), and VRPF (green-plain).

#### 7.2.1 Simulator Sickness Questionnaire (SSQ)

Results of the SSQ in terms of sub-scales and total are summarized in Figure 12. Because of the non-normality of the data, we performed a series of Friedman tests, which showed that there was a statistically significant difference in Oculomotor and Dizziness, $\chi^2(2) = 14.292$, $p = 0.001$ and $\chi^2(2) = 8.667$, $p = 0.013$ respectively. On the other hand, Nausea and Total did not appear to differentiate significantly according to the test $\chi^2(2) = 1.755$, $p = 0.416$ and $\chi^2(2) = 5.729$, $p = 0.057$, respectively.

During the post-hoc analysis with Wilcoxon signed-rank tests, we observed the Median (IQR) Oculomotor levels for the VRPF, BM, and SM were 0, 2, and 1, respectively. After correction there were significant differences between VRPF and SM ($Z = -2.365$, $p = .018$) and VRPF and BM ($Z = -3.169$, $p = .002$), but not between BM and SM ($Z = -1.512$, $p = .131$). We further observed significances in Dizziness, VRPF and SM ($Z = -2.337$, $p = .019$) and VRPF and BM ($Z = -2.209$, $p = .027$), but not for BM and SM ($Z = -0.686$, $p = .492$). Median (IQR) Oculomotor levels for the VRPF, BM, and SM were 3, 2 and 1.

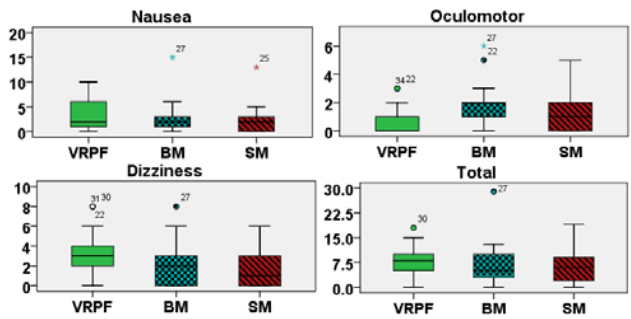

Figure 12: The results of SSQ for four sub-scales in Experiment C. VRPF generated more Dizziness than BM and SM, but less in Oculomotor distress.

### 7.2.2 User Performance

Figure 13 shows the data for Score, Time, and Accuracy across the three versions. Like Experiment A, Shapiro-Wilk revealed a lack of normality for Score, but Time and Accuracy data were normal. The RM-ANOVA suggested that there were no significant main differences in Accuracy ($F_{2, 34} = .677$, $p = .515$) among the three versions. No significant relations were found for Time ($F_{2, 34} = 0.26$, $p = .974$) either. On average, each condition lasted 307 seconds (5.1 minutes).

Unlike in Experiment A the Friedman test revealed a significant difference for Score $\chi^2(2) = 8.121$, $p = 0.017$. The Wilcoxon test revealed that the difference was significative between BM and SM ($Z = -2,182$, $p = .029$) and between BM and VRPF ($Z = -2,535$, $p = .011$). Their medians were BM = 74.9, SM = 69.8, and VRPF = 61.2.

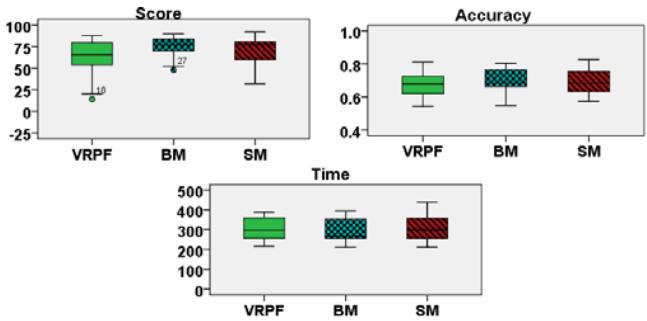

Figure 13: The results of Score, Accuracy, and Time in Experiment C. BM presented a higher Score than the other versions.

### 7.2.3 Subjective Immersion

Among the three studied versions, Overall Immersion did not present significant differences according to the RM-ANOVA ($F_{2, 34} = .535$, $p = .591$). Figure 14 shows the data for Overall Immersion across the versions.

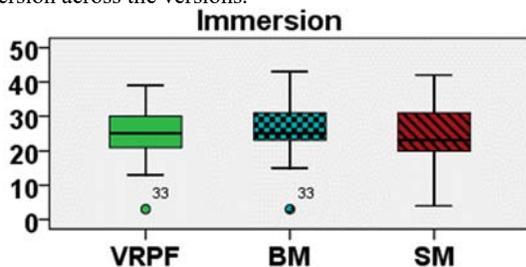

Figure 14: The results of Overall Immersion for Experiment C. Overall Immersion across versions did not differ significantly.

## 7.3 Discussion

In this experiment, we observed that different variations of PF resulted in similar levels of user performance and immersion. BM and SM presented virtually no difference except in the oculomotor component of SS, which was also statistically insignificant, suggesting that our results are not the fruit of the FOV, unlike the work of [30] that used some kind of visual technique to hide only the outer region without a 2D effect.

Both SM and BM presented reduced levels of Dizziness when compared to VRPF, but also presented increased levels of Oculomotor discomfort. The darker outer region could have allowed for the eyes to rest. However, this was a liability when associated with the fast movement of the controller.

From the results, it seems that it can be advantageous to keep some level of the PF view besides the PF view when using a controller because it might aid in the Dizziness factor. If the oculomotor strain is the most significant concern, opting for the darker outer layer can be of more considerable aid, similar to [30].

Since there was no notable difference among these techniques in this experiment, it can be concluded that all of them are somewhat viable for FPS games, even considering the SS. However, considering the results from Experiment B, it may be useful to investigate SM further, since it generated less Nausea than VRPF and lower levels of Dizziness as well.

## 8 BETWEEN-SUBJECTS ANALYSIS (EXPERIMENTS A AND B)

The goal of this analysis was to compare how vital the controller is for user performance and immersion when playing an FPS game in VR HMD. We compared the keyboard against the recommended VR controller. We were interested in exploring if the controller would influence the dependent variables. Therefore, in this analysis, we compared four conditions based on two inputs and two views: VRPF-Keyboard (1), VRPF-Controller (2), RVR-Keyboard (3), and RVR-Controller (4).

We compared the data of participants from Experiments A and B. However, to avoid confounding factors, we only compared the data from the first version played by each participant. As such, we have a total of 24 participants, 6 played VRPF and 6 RVR using a keyboard and 12 participants who did the same using the Oculus Touch Controller. We performed two between-subjects analyses using the data, one comparing the differences between keyboard and controller in VR regardless of the view and one only between subjects who had experienced the same view.

### 8.1 Results

We employed a similar procedure to analyze the data as the first experiment, first checking for normality, followed by investigating significant differences with ANOVA for parametric data; otherwise, Kruskal-Wallis and Mann-Whitney was used. We used the following colors and patterns to represent the four versions in the plots: 1 (light green – vertical stripes), 2 (dark green - diagonal stripes), 3 (light red – vertical stripes), and 4 (dark red – diagonal stripes).

#### 8.1.1 Simulator Sickness Questionnaire (SSQ)

Figure 15 shows the results of the SSQ in terms of sub-scales and Total. Because the data were not normally distributed, the Mann-Whitney test was used and showed that there was a significant difference in Nausea (Mann-Whitney $U = 24.5$, $P < 0.05$), Oculomotor ($U = 28.0$, $P < 0.05$) and Total ($U = 22.0$, $P < 0.05$) based on the input alone. No significant difference was found in Dizziness ($U = 66.0$, $P > 0.05$).

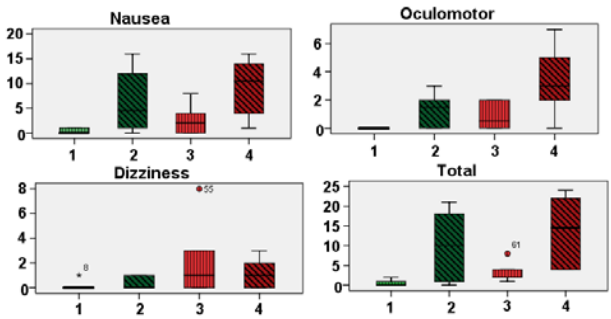

Figure 15: The results of SSQ concerning four sub-scales in the between-subjects analysis. The input method was hugely influential.

Comparing inputs within a view Mann-Whitney tests also presented significant differences in Nausea for both RVR (U = 5.0, P < 0.05), and VRPF (U = 6.0, P < 0.05); Oculomotor for both RVR (U = 5.5, P < 0.05), and VRPF (U = 6.0, P < 0.05). Total had similar results for RVR (U = 3.0, P < 0.05) and VRPF (U = 5.5, P < 0.05).

#### 8.1.2 User Performance

When analyzing input alone, the Kruskal-Wallis H test did not show a statistically significant difference in Accuracy ($\chi^2(2)$ = .654, p = .419), with a mean rank of 13.67 for Controller and 11.33 for Keyboard. However, there was a significant difference in Score ($\chi^2(2)$ = 6.321, p = 0.012), with a mean rank of 8.88 for Controller and 16.13 for Keyboard.

Results from a one-way ANOVA showed no significant effect for Time (F1, 22 = 0.013, p = .911); on average each condition lasted 293 seconds (4.9 minutes). The medians can be seen in Figure 16.

The input conditions presented differences in Score when the players were using the VRPF (U = 3.5, P < 0.05 two-tailed) with a mean rank of 8.92 for 1 and 4.08 for 2.

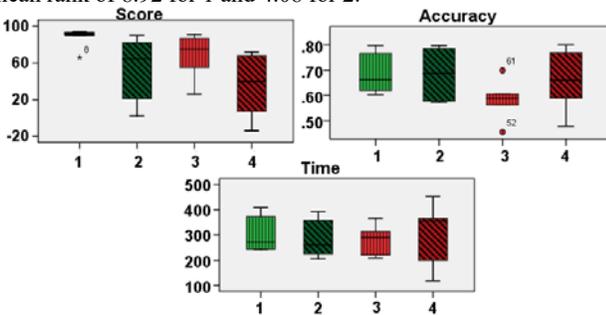

Figure 16: The results of Score, Accuracy, and Time in the between-subjects analysis. The keyboard yielded better results in Score; Accuracy relied on the input method matching the view.

#### 8.1.3 Subjective Immersion

When comparing all views, the Kruskal-Wallis H test did not show a statistically significant difference ($\chi^2(2)$ = 4.984, p = 0.173), with a mean rank of 16.75, 7.83, 13.58 and 11.83 for conditions 1-4, respectively. When checking without considering the 2D/3D aspect, the difference was still non-existent (Mann-Whitney U = 40, P > 0.05). However, when comparing input in VRPF, there a difference in immersion (U = 5.0, P < 0.05); details can be seen in Figure 17.

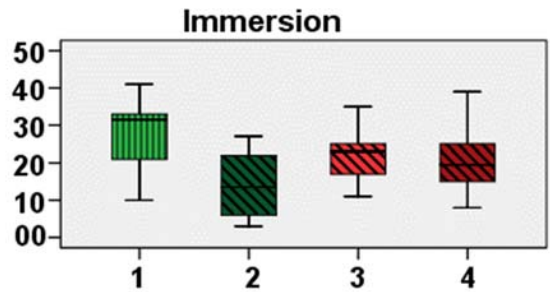

Figure 17: The results of Overall Immersion for the between-subjects analysis. Overall Immersion across versions did not differ significantly.

### 8.2 Discussion

Our results suggest that the kind of controller used can be the most influential aspect of FPS when it comes to SSQ Nausea and Total. The keyboard generated lower levels of simulator sickness when compared to the regular Oculus controller when playing the FPS VR game. In addition, the Score measure also presented an improvement when using the keyboard combined with our VRPF, which suggests that the use of controllers might not be beneficial for people who are using a 2D view. Similarly, keyboards might be less precise for those who are using a 3D view.

Interestingly, for Overall Immersion, our results did not show a significant difference when using the keyboard in RVR. These results somewhat contradict findings from [3]–[5]. The cause might be simply due to how the other studies were conducted. For example, in [3]–[5], the authors did not count for the cofound variable of view, which the results in Experiments A and B show to affect immersion. Moreover, FPS games might be less susceptible to variations in immersion, as suggested in previous studies (e.g., [17]). *RQ2* is answered with mixed results. Our results support that subjective feelings of physiological background are more affected by the controller than by the view. However, our results suggest that the view can be highly impactful, even if it has an unnatural synergy with the controller.

Our results strongly suggest that using a Keyboard can bring benefits for the overall VR experience. Our results do not support *HA1*—playing with the keyboard is not detrimental to the immersive experience; it yielded a good score performance and was superior in immersion when using a 2D view. Our results support *HA2*—the 2D view can be immersive, and it lets the players focus solely on the game; as such, overall, it is not detrimental to the immersive experience. For FPS games, the overall superior playability and the possibility of spending extended periods in the games might make the keyboard a preferred choice. For *RQ1*, our results indicate that the controller seems to have a more significant part in the performance than the view does. However, for the maximum yield of each controller type, it must be harmonized with the adequate type of view, because score performance was the best when the view matched its expected controller type. Based on the answers to *RQ1* and *RQ2*, the answer to *RQ3* appears to be mixed: the stereoscopic 3D view alone is not a significant influence; however, its combination with the control method can be highly significant.

## 9 RECOMMENDATIONS AND POTENTIAL APPLICATIONS OF PF

Based on our findings, we recommend the use of the VRPF version of our PF associated with a keyboard for VR FPS games. This combination seems to lead players to enjoy an immersive experience while obtaining their best performance scores. Besides, PF players appear to be able to play the game longer because, though they are fully immersed in VR, they are likely to feel less sick.

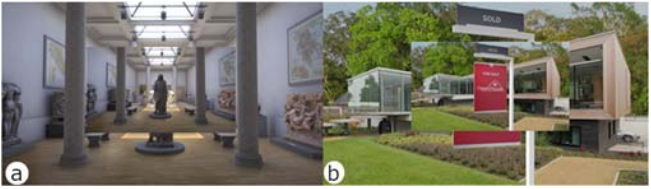

Figure 18: a) Demo of a museum visit using PlaneFrame Mask. The user can appreciate the exposition and travel/look around for a longer time, which is adequate for such a learning environment that requires attention to detail as simulator sickness can be reduced. b) Demo of an open house or architectural 3D VR environment. The user can have a complete understanding of the space, due to the linear movement provided by the technique and without having the vision occluded by a gaze point.

In an environment that does not allow the use of a keyboard, we recommend the use of an SM (see Experiment C) because it keeps most of the SS mitigation benefits provided by VRPF, while still letting players immersed and allowing them to achieve high Accuracy.

PF can be inserted in many games. FPS and Racing game characters can justify the use of the technique through the use of protective gear. In horror games, it can be a device to bring focus to a specific scene. In addition, PF could also be used in applications outside of games that do not require fast responses. For example, educational environments such as museums (see Figure 18a) and other forms of VR education and training environments can make use of the technique. It allows students to be immersed for extended periods or let them move more and explore further the environment. Extensive exploration is also useful for other kinds of applications, such as an open-house VE (see Figure 18b). In this kind of scenario, the user likely wants and needs a clear understanding of the space. Such clear understanding could be challenging to achieve when users feel sick and want to disengage from the system because further exposition will have adverse effects after every movement.

## 10 LIMITATIONS AND FUTURE WORK

We studied three variations of PlaneFrame (PF). Due to the continuous nature of the mask FoV, we did not explore all possible sizes for the frame. Nevertheless, given that the two sizes we explored did not lead to significant differences in the results, it is likely the case for other possible sizes. Further explorations are needed to ascertain if this is the case.

We explored the use of PF in one specific kind of game (FPS). We chose FPS games because of the fast-moving nature and constant change of viewpoints; they are often regarded as ideal environments to investigate SS mitigation techniques (see [41], [51]). These techniques have also been used outside of the context of VR HMD [51]–[54]. Our results could be applied to other VR applications with less demanding navigation tasks. Nonetheless, it will be useful to explore the applicability and use of PF in other types of applications. For future work, we would like to explore if we can achieve the same results in other types of games.

As the main aim of our study was to identify the effect of view mode and controller type on people's level of immersion, SS, and performance for VR FPS, we have been able to develop a viewing technique that led to improved results in VR FPS. In the future, it will be useful to explore the combination of PF with other types of devices and techniques that also aim to reduce SS [36], [43]. We can evaluate users' preferences and assess their combined effect.

Further, we can analyze other tasks like locating objects in 3D VE and spatial memory recall of the location of items in these environments [24].

## 11 CONCLUSION

This paper presents an in-depth exploration of the effect of different view dimensionalities and input methods in virtual reality (VR) gaming, with a focus on first-person shooter games. We studied how such interactions affect immersion, accuracy, and simulator sickness. To do so, we developed a concept technique PlaneFrame (PF) to study 2D/3D views in VR. In the first of three studies, we investigated the 2D/3D views and display using a keyboard in a first-person shooter game. The results showed that with PF users achieved better performance and had lower SS than playing the game in the original, unmodified VR display. In our second experiment, we compared PF with other VR conditions using a controller (gaze pointing and original, unmodified VR display). In our last experiment, we presented and evaluated three variations of PF. In our final analysis, we observed that even though the view did play a factor in all the components, the controller type was more influential.

Our results suggest that PF is useful in reducing the levels of SS in VR while maintaining and possibly boosting players' performance. It is a novel technique that is simple to implement with nearly no associated computing overhead. It also does not require additional external devices, is not intrusive, and is relatively easy for users to adapt to and use it. Keyboards are preferred when there is no need for fast turning, and 2D is acceptable. We conclude that further investigations should focus on controllers rather than viewing because of its considerably stronger influence in the diverse aspects of VR.

## ACKNOWLEDGMENTS

We would like to thank the participants for their time. This research was partially funded by AI University Research Centre at Xi'an Jiaotong-Liverpool University (XJTLU), XJTLU Key Program Special Fund (KSF-A-03 and KSF-02) and XJTLU Research Development Fund.